\documentclass[%
aip,
sd,%
amsmath,amssymb,
reprint,%
author-numerical,%
]{revtex4-1}

\usepackage{graphicx}
\usepackage{dcolumn}
\usepackage{bm}
\usepackage[dvipsnames]{xcolor}
\usepackage[normalem]{ulem}
\usepackage[title]{appendix}
\usepackage{comment}
\usepackage{float}

\graphicspath{{./figures/}}
\begin{document}

\title{ Reservoir Computing with Random and Optimized Time-Shifts }

\author{Enrico Del Frate}
	\affiliation{ 
		Mechanical Engineering Department, University of New Mexico, Albuquerque, NM 87131}

\author{Afroza Shirin}
\affiliation{ 
	Mechanical Engineering Department, University of New Mexico, Albuquerque, NM 87131}

\author{Francesco Sorrentino}
	\email{fsorrent@unm.edu}
	\affiliation{
		Mechanical Engineering Department, University of New Mexico, Albuquerque, NM 87131}

\begin{abstract}
	We investigate the effects of application of random time-shifts to the readouts of a reservoir computer in terms of both 
	accuracy (training error) and performance (testing error.) 
	For different choices of the reservoir parameters and different `tasks', we observe
	a substantial improvement in both accuracy and performance. We then develop a simple but effective 
	technique to optimize the choice of the time-shifts, which we successfully test in numerical experiments.
\end{abstract}

\maketitle

\textbf{We study how the accuracy and performance of a reservoir computer (RC) can be enhanced by application of 
\textcolor{black}{different}
time-shifts to the RC readouts. Our numerical analysis shows an improvement for different parameters of the RC dynamics and for different `tasks', such as reconstructing the attractors of several 
chaotic dynamical systems. For certain tasks, the attained improvement is of several orders of magnitude.}

\section{Introduction}

A reservoir computer (RC) is a complex nonlinear dynamical system that is used for processing and analyzing empirical data, see e.g. \cite{jaeger2001echo, schrauwen2007overview, natschlager2002liquid, maass2002real, martinenghi2012photonic,brunner2013parallel,nakajima2015information, hermans2015photonic,vinckier2015high,duport2016fully,larger2017high}, 
modeling of complex dynamical systems \cite{suykens2012artificial}, 
speech recognition \cite{crutchfield2010introduction},  learning of context free and context sensitive languages \cite{rodriguez2001simple, gers2001lstm},
the reconstruction and prediction of chaotic attractors \cite{lu2018attractor,antonik2018using, jaeger2004harnessing,pathak2017using,pathak2018model},
image recognition \cite{jalalvand2018application},
control of robotic systems \cite{graves2004biologically, robinson1994application,lukovsevivcius2012reservoir}, predicting catastrophic critical transitions \cite{kong2021machine} and amplitude death in oscillating systems \cite{xiao2021predicting}. 
A typical RC consists of a set of nodes coupled together to form a network. Each node of the RC evolves in time in response to an input signal that is fed into the reservoir. An output signal is then generated from the time evolutions of the RC nodes.
In an RC, the output connections (those that connect the RC nodes to the output) are trained to produce a best fit between the output signal and a training signal related to the original input signal. On the other hand, the connections between the nodes of the reservoir are constant parameters of the system. As a result, RCs are easier to analyze than other machine learning tools for which all the connections are typically  trained.



The performance of an RC depends on variety of factors such as nonlinearity of the nodal dynamics \cite{dambre2012information,shirin2019stability}, network topology, sparsity of the connections and the presence of network symmetries \cite{carroll2019network}, input signal and the dynamic range of the input signals \cite{verstraeten2009quantification} and  time-delay structure of the RC \cite{martinenghi2012photonic,hermans2015photonic,larger2017high}. Experimental realizations of RCs have been proposed in \cite{brunner2013parallel,larger2017high,vinckier2015high,sheldon2020computational}, among other papers. Recent work  has analyzed linear RCs \cite{boyd1985fading},\cite{bollt2021explaining} and pointed out a connection with the theory of dynamic mode decomposition \cite{schmid2010dynamic}.

{
A universal approximation theorem and its application to reservoir computers with stochastic inputs has been presented in \cite{gonon2019reservoir}. In \cite{bollt2021explaining,gauthier2021next} it has been shown that a reservoir computer can perform as a universal representor of a dynamical system. Other papers have shown dramatic effects of tuning several parameters and hyper-parameters of RCs, see e.g.,
\cite{carroll2019network,carroll2020dimension,carroll2020reservoir}. In particular, Ref.\ \cite{carroll2019network} investigated the effects of the sign of the weights associated with the network edges as well as the symmetries of the network topology and shown that network symmetries are usually undesirable in terms of the performance of RCs.}

 In this paper, we focus on the effects of time-shifts applied to the readouts of an RC.
 In the literature there has been documented improvements in RC performance by applying a single time-shift to all nodes\cite{stelzer2020performance}, however, applying different time-shifts to individual nodes is new.
  References \cite{martinenghi2012photonic,larger2017high} focused on the case that the time evolution of the RC obeys a delay differential equation. This is different from what we do here where the RC dynamics is described by an ordinary differential equation; once the RC dynamics is computed, different time-shifts are applied to the individual RC readouts. 
  
 In the first part of this paper,
 random time-shifts are applied at the readout of each node of an RC, which produces an improvement in both accuracy and performance.  In the second part of this paper, a simple optimization technique is implemented to optimize the time-shift at each node in order to further improve the accuracy and performance of an RC.  Optimizing the hyperparameters of an RC is often done, but optimizing the time-shift  at each node of an RC is more difficult due to the high-dimensional parameter space. {Our numerical analysis shows that for different parameters of the RC dynamics, and for different `tasks', an RC with time-shifts provides an increase in accuracy and performance.  }

\section{Methods}
\subsection{Reservoir Dynamics}
We consider an RC modeled by the following nonlinear dynamical equations in continuous time \cite{griffith2019forecasting},
\begin{equation} \label{eq:nonlinear}
    	\dot{\textbf{r}}(t) = \gamma\left[-\textbf{r}(t) + \tanh(\epsilon A \textbf{r}(t) + s(t) \textbf{w})\right],
\end{equation}
where $\textbf{r}(t)$ is the $N$-dimensional state vector of the reservoir and $s(t)$ is the input signal, the $N$-dimensional symmetric adjacency matrix $A=\{A_{ij}\}$ describes the connectivity between the $N$ nodes of the network and the $N$-dimensional vector $\textbf{w}$ are the weights by which the input signal is multiplied. In what follows we refer to the time evolutions $\textbf{r}(t)$ as the readouts of the RC.
In this paper we set $N=100$.  
%
%
The adjacency matrix $A$ is constructed such that it is symmetric and its off-diagonal entries are uniformly drawn at random from the interval $[0,1]$. The entries on the main diagonal of the matrix $A$ are all set to be equal to $\beta<0$, where the scalar $\beta$ is negative enough to ensure that all the  eigenvalues of the matrix $A$ are less than $0$. The variable parameter $\epsilon$ is used to tune the spectral radius of the matrix $A$. The entries of the vector $\textbf{w}$ are all chosen to be 1.  The variable parameter $\gamma>0$ determines the time-scale on which the RC dynamics evolves.\\

The underlying process we want to model may evolve in time based on a set of deterministic (chaotic) equations, such as the equations of the Lorenz chaotic system, in the variables $x(t), y(t), z(t)$ (See Eq.\ \eqref{Lorenz system}.)
One task that can be given to the RC is to reconstruct the time evolution of the training signal, e.g. $y(t)$ from knowledge of the input signal, e.g., $x(t)$. In this paper we will consider several similar tasks for which the time series are generated by various chaotic systems. An example of chaotic higher-dimensional system we will use in this paper is the Lorenz96 system  \cite{lorenz1996predictability}. For this system, we see particularly strong benefits of introducing the time-shifts.  

\subsection{Training and Testing Error of The Reservoir Computer}
In order to examine the accuracy of the RC relative to the dynamical system it is modeling, we need to quantify how well the reservoir is able to reproduce the training signal $g(t)$ from knowledge of the input signals $s(t)$. An RC driven by the input signal has three phases: the transient phase which is from $t_0 = 0$ to $t_1$,  the training phase which is from $t_1$ to $t_2$, and the testing phase which is from $t_2$ to $t_3$. 

During the training phase $[t_1 \quad t_2]$ the readouts from each node are recorded, discretized, and combined in a $ T \times (N+1)$ matrix,
\begin{equation}
\Omega =
	\left[\begin{array}{ccccc}
	    r_1(1) & r_2(1) & ... & r_N(1) & 1 \\
	    r_1(2) & r_2(2) & ... & r_N(2) & 1\\
	    \vdots & \vdots & \vdots &  \vdots  & \vdots\\
	    r_1(T) & r_2(T) & ... & r_N(T) & 1\\
	\end{array}\right]
\end{equation}
Here, $N$ is the number of nodes in the RC and $T$ is the number of time-steps recorded in the interval $[t_1 \quad t_2]$. We add a  column whose entries are all ones to account for any constant offset. The fit $\mathbf h = [h(1),h(2),...,h(T)]$ to the training signal  $\mathbf g = [g(1),g(2),...,g(T)]$ is equal to,
\begin{equation}
	h(t)=\sum_{i=1}^N \kappa_i r_i(t)+\kappa_{N+1}
\end{equation}
(or, equivalently, in vectorial form
$\mathbf h = \Omega \pmb{\kappa}$),
where  the vector $\pmb{\kappa} = [\kappa_1,\kappa_2,...,\kappa_{N+1}]$, which contains a set of unknown coefficients to be determined. The weight vector $\pmb{\kappa}$ is obtained by minimizing the linear least square fit problem,
\begin{equation} 
	\min_{\pmb{\kappa}} \quad ||\Omega\pmb{\kappa} -\mathbf{g}||_2^2.
\end{equation}
The analytic solution of the problem is given by,
\begin{equation} \label{kappa}
	\pmb{\kappa}= \left(\Omega^T\Omega\right)^{-1} \Omega^T {\mathbf{g}},
\end{equation}
When the matrix $\Omega$ is super-collinear (columns are highly linearly dependent or $T>> N$),  the inverse of the matrix $\Omega^T\Omega$ is difficult to compute numerically. To avoid this, we estimate the weight vector $\pmb{\kappa}$ as a solution of the linear least square fit problem with ridge regression,
\begin{equation} 
	\min_{\pmb{\kappa}} \quad ||\Omega\pmb{\kappa} -\mathbf{g}||_2^2 + \eta ||\pmb{\kappa}||_2^2,
\end{equation}
where $\eta>0$ is a small positive number. The solution of the above problem can be computed as,
\begin{equation} 
	\pmb{\kappa}_{\text{ridge}} = \left(\Omega^T\Omega + \eta I\right)^{-1} \Omega^T {\mathbf{g}}.
\end{equation}
Here $I$ is the identity matrix of size $N$. 
From this, we can compute the training error,
\begin{equation}\label{eq:train}
	\Delta_{tr} = \frac{\langle \Omega \pmb{\kappa}_{\text{ridge}} - \mathbf g \rangle}{\langle \mathbf g \rangle},
\end{equation}
where the notation 
$\langle \mathbf X \rangle = \sqrt{\frac{1}{T} \sum_{i=1}^T (X(i) - \mu)^2}$
for $\mathbf X$ any $T$-dimensional vector and $\mu = \frac{1}{T} \sum_{i=1}^T X(i)$. 

A fundamental measure of the performance of an RC is the testing error. The testing error is defined as,
\begin{equation}\label{eq:test}
	\Delta_{ts} = \frac{\langle \tilde{\Omega}  \pmb{\kappa}_{\text{ridge}} - \mathbf {\tilde{g}} \rangle}{\langle \mathbf{ \tilde{g}}  \rangle},
\end{equation}
where  $\mathbf{ \tilde{g}}(t)$ is the testing signal we want to estimate, $ \tilde{\Omega}$ contains the time evolutions from the RC over the time interval $[t_2 \quad t_3]$, and $\pmb{\kappa}_{\text{ridge}}$ is the same coefficient vector we found in the training phase.

\subsection{Initial setting of the reservoir parameters $\gamma$ and $\epsilon$}

{In Secs.\ III and IV we will consider the effects of application of time shifts to a \textcolor{black}{well performing} Reservoir Computer, meaning that the RC has been preliminarily
optimized based on current state-of-the-art practices. However, we stress out that we have seen similar improvements when the RC is not optimized.

Our preliminary optimization consists of two steps: (i) optimization in the coefficient $\gamma$ and (ii) optimization in the coefficient $\epsilon$, see Eq.\ (1). We first discuss (i) and then (ii). For each one of the tasks, in order to pick a best value of $\gamma$, we set $\epsilon=1$ and compute the training error $\Delta_{tr}$ as a function of $\gamma$ in the interval $[0,5]$; we then pick the value of $\gamma$ that minimizes $\Delta_{tr}$. We keep $\gamma$ at the selected value and investigate the effect of varying $\epsilon$. In order to pick the best value of $\epsilon$ we compute the memory capacity defined as follows \cite{jaeger2001short},
\begin{equation}
MC=  \sum_{\tau=1}^{\infty} MC_\tau, 
\end{equation}
where
\begin{equation}
MC_{\tau} =  \frac{\text{cov}^2(x(t-\tau),h(t))}{\text{var}(x(t))\text{var}(h(t))}, \quad \tau \in \mathbb{N}
\end{equation}
and select the value of $\epsilon$ that maximizes the MC.}

This procedure (optimization in $\gamma$ followed by optimization in $\epsilon$) is illustrated in Fig.\ 1 for the case of the Lorenz96 system \cite{lorenz1996predictability}, see also Eq.\ \eqref{L96} below. From (A) we see that the value of $\gamma$ that minimizes the training error is approximately equal to $0.9$. We then fix $\gamma=0.9$ and vary $\epsilon$, which is shown in (C). For completeness, the testing error is shown in (B). We see that the value of $\epsilon$ that maximizes the memory capacity is approximately equal to $0.8$. Analogous procedures are implemented for the cases of the Lorenz system and of the Hindmarsh-Rose system, which we discuss later in Sec.\ III. For the Lorenz system we obtain $\gamma=1.65$ and $\epsilon=1$. For the Hindmarsh-Rose system we obtain $\gamma=0.9$ and $\epsilon=0.8$.

\begin{widetext}
    \begin{minipage}{\linewidth}
       	\begin{figure}[H]
			\centering
    	\includegraphics[width=0.60\textwidth]{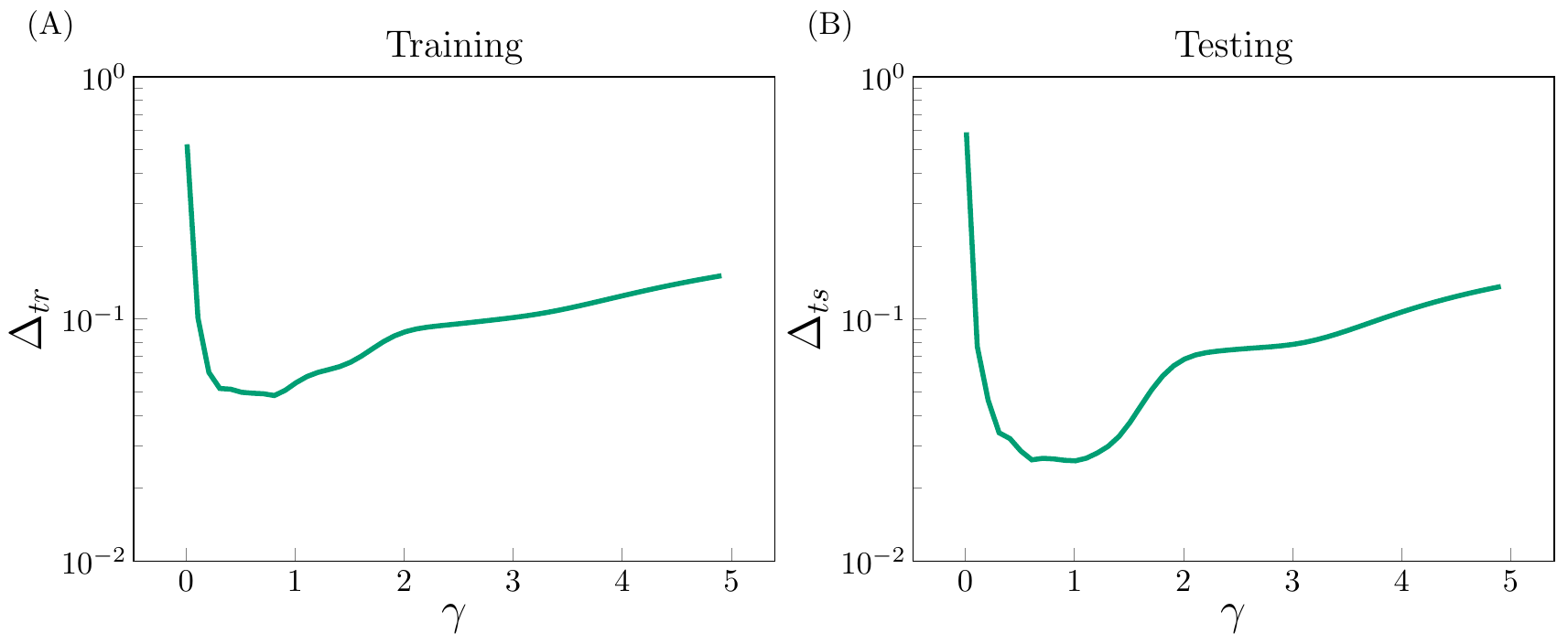}
\includegraphics[width=0.38\textwidth]{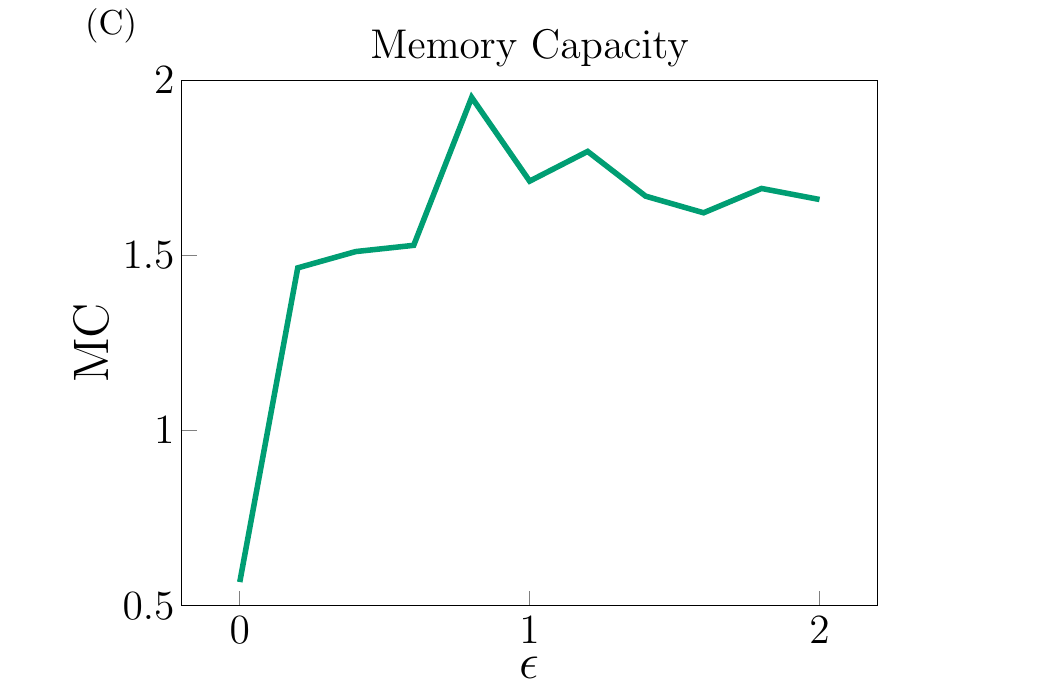}
			\caption{Lorenz96 attractor. (A) Plot of the training error $\Delta_{tr}$ vs $\gamma$. The value of $\gamma$ that minimizes the training error is approximately equal to $0.9$. (B) Testing error vs $\gamma$, with a trend similar to that seen in (A). (C) Memory Capacity vs $\epsilon$ with $\gamma$ set equal to $0.9$. We find that the value of $\epsilon$ that maximizes the memory capacity is approximately equal to $0.8$.  }\label{fig:Lorenz96 Gamma opt}
		\end{figure}     
    \end{minipage}
\end{widetext}

\section{Application of Random Time-Shifts}
In this section we describe application of time-shifts to the individual readouts of a RC. 
We will see that introduction of these time-shifts is beneficial even when these are randomly chosen. Optimized time-shifts are considered in Sec.\ IV. Here our choice of random time-shifts is consistent with the choice of a random topology for the connectivity of the RC network, which is commonly assumed in the literature (see also our construction of the adjacency matrix $A$ in section II.) This is typically done to show that RCs can be effective independent of the details of their implementation. We remove the assumption of randomly chosen time-shifts in Sec.\ IV.

For each individual task, we compute the timescale $\bar{\tau}$ of each individual oscillator system, defined as the time at which the system  autocorrelation function decays to one half of its value at time zero. For the Lorenz system we find $\bar{\tau}=0.3$; for the Hindmarsh-Rose system  we find  $\bar{\tau}=0.46$; and  for the Lorenz96 system we find  $\bar{\tau}=0.19$. Subsequently, for each task, the individual time shifts $\tau_i$ are taken to be uniformly distributed random numbers in the interval $[0,\alpha \bar{\tau}]$, where $\alpha$ is a tunable parameter. 

\color{black}

Finally, the reservoir readout at node $i$ is shifted $r_i \mapsto r_i(t-\tau_i)$. The motivation for application of the time shifts is the observation that under general conditions the RC readouts $r_1(t), r_2(t), ...$ appear to be `synchronized' \cite{nathe2021reservoir},  which significantly reduces the ability of fitting the training signal. By introducing time shifts, this synchronization can be broken.  

The fit signal $\textbf{h}(t)$ is written as a linear combination of the individual readouts,
\begin{equation}\label{eq:fit_delay}
	\textbf{h} (t) = \Omega_{\text{delay}}\pmb{\kappa}_{\text{ridge}}
\end{equation}
where $\pmb{\kappa}_{\text{ridge}}$ in this case is computed as,
\begin{equation} 
	\pmb{\kappa}_{\text{ridge}} = \left(\Omega_{\text{delay}}^T\Omega_{\text{delay}} + \eta I\right)^{-1} \Omega_{\text{delay}}^T {\mathbf{g}}
\end{equation}
and $\Omega_{\text{delay}}$ is computed as,
\begin{equation}\label{eq:Omega_delay}
	\Omega_{\text{delay}} =
	\left[\begin{array}{ccccc}
		r_1(1-\tau_1) & r_2(1-\tau_2) & ... & r_N(1-\tau_N) & 1 \\
		r_1(2-\tau_1) & r_2(2-\tau_2) & ... & r_N(2-\tau_N) & 1\\
		\vdots & \vdots & \vdots &  \vdots  & \vdots\\
		r_1(T-\tau_1) & r_2(T-\tau_2) & ... & r_N(T-\tau_N) & 1\\
	\end{array}\right]
\end{equation}
In the testing phase the same times-shifts used in the training phase, are applied to compute $\tilde\Omega_{\text{delay}}$. Then the training and testing errors are computed by using Eq.\ \eqref{eq:train} and \eqref{eq:test}, respectively.

In the rest of this section, we provide evidence of the strong benefits of applying randomly chosen time shifts, rather than presenting a principle way of selecting them. In Section IV we present an optimization approach that can be used to guide the selection of the time-shifts.

\begin{widetext}
	\begin{minipage}{\linewidth}
		\begin{figure}[H]
			\centering
			\includegraphics[width=0.99\textwidth]{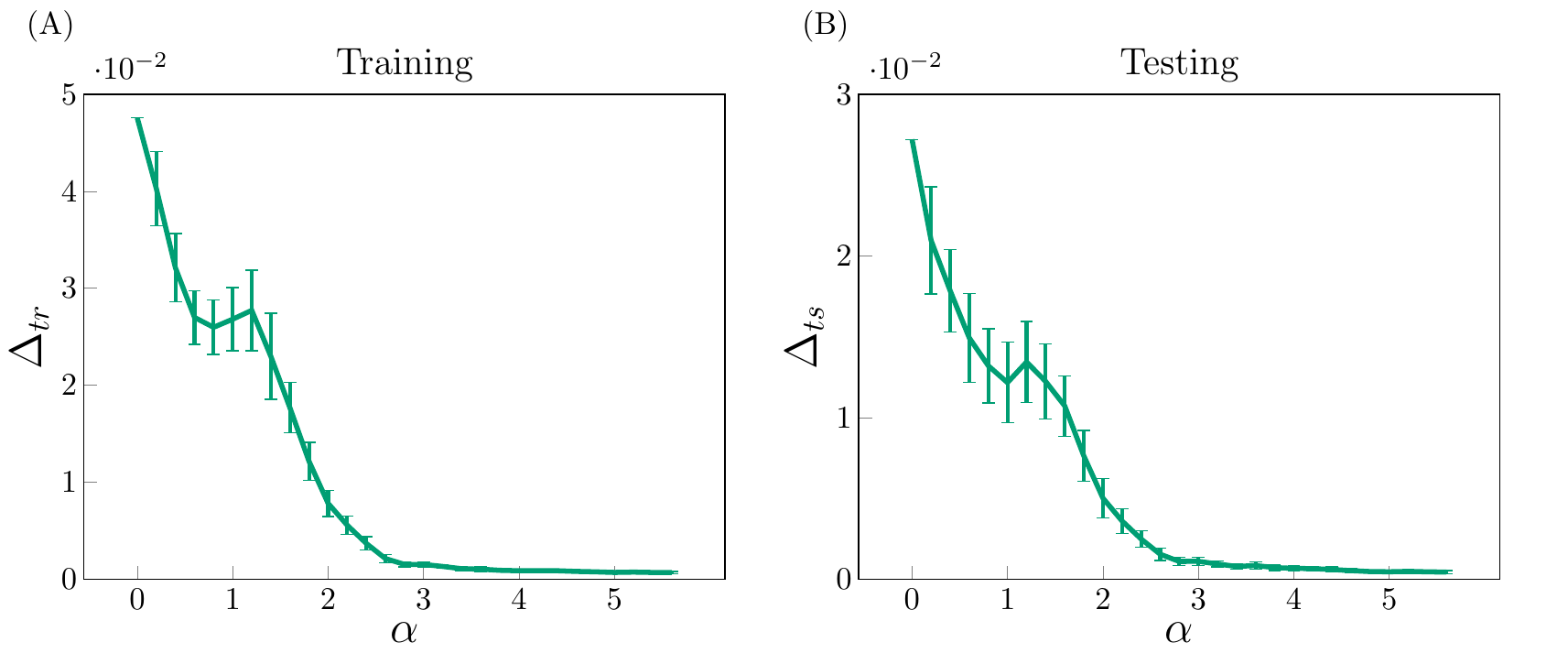}
			\caption{Lorenz96 system.  The training error (A) and testing error (B) vs $\alpha$. The parameter $\alpha$ controls the interval over which the random time-shifts are taken $[0,\bar{\tau}\alpha]$, where $\bar{\tau}$ is the characteristic time-scale of the Lorenz task and was found to be 0.19. {Error bars indicate the standard deviation over 50 iterations where each iteration corresponds to a different selection of the random time-shifts $\tau_i$, $\gamma=0.9$, $\epsilon=0.8$.}} \label{L96R}
		\end{figure}    
	\end{minipage}
\end{widetext}

\begin{widetext}
	\begin{minipage}{\linewidth}
		\begin{figure}[H]
			\centering
    	\includegraphics[width=0.99\textwidth]{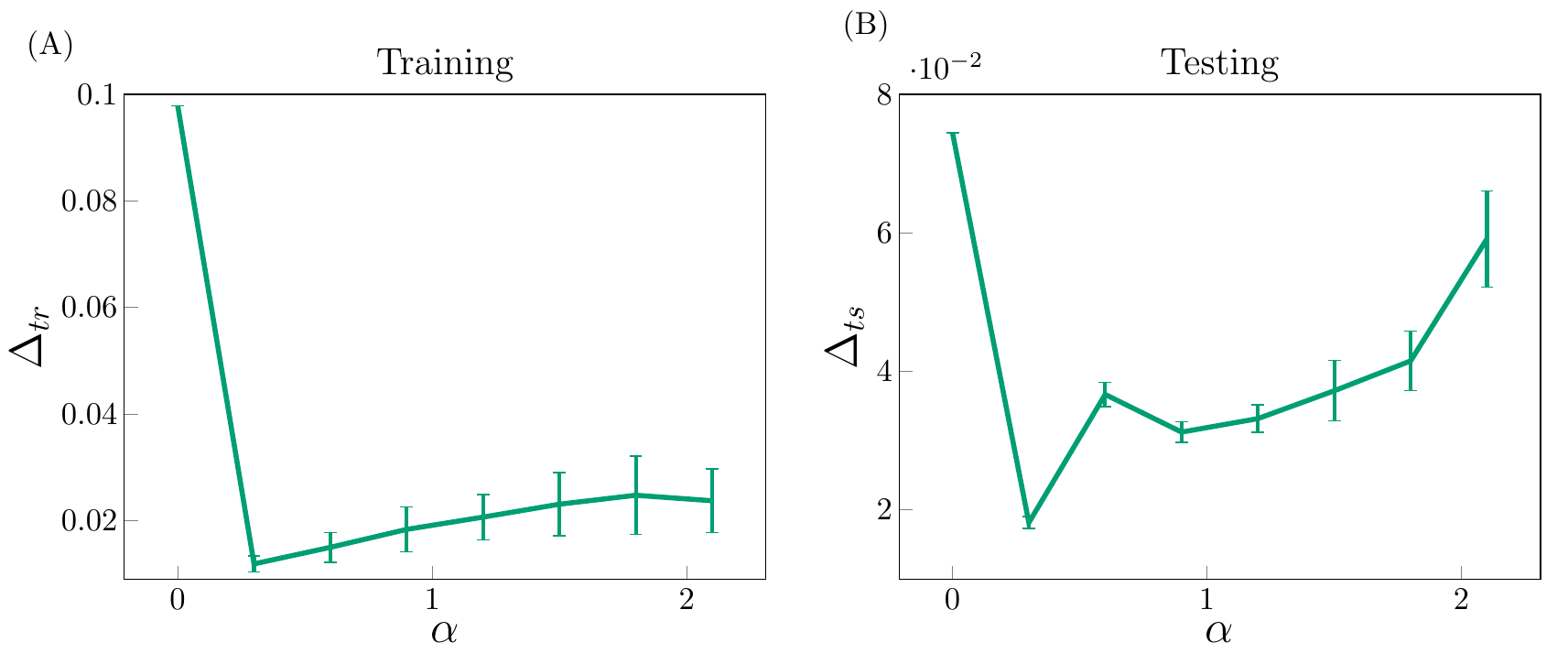}
			\caption{Lorenz attractor. The training error (A) and testing error (B) vs $\alpha$. The parameter $\alpha$ controls the interval over which the random time-shifts are taken $[0,\bar{\tau}\alpha]$, where $\bar{\tau}$ is the characteristic time-scale of the Lorenz task and was found to be 0.3. {Error bars indicate the standard deviation over 50 iterations where each iteration corresponds to a different selection of the random time-shifts $\tau_i$, $\gamma=1.3$, $\epsilon=2$.} }\label{LR}
		\end{figure}    
	\end{minipage}
\end{widetext}

\begin{widetext}
	\begin{minipage}{\linewidth}
		\begin{figure}[H]
			\centering
			\includegraphics[width=0.99\textwidth]{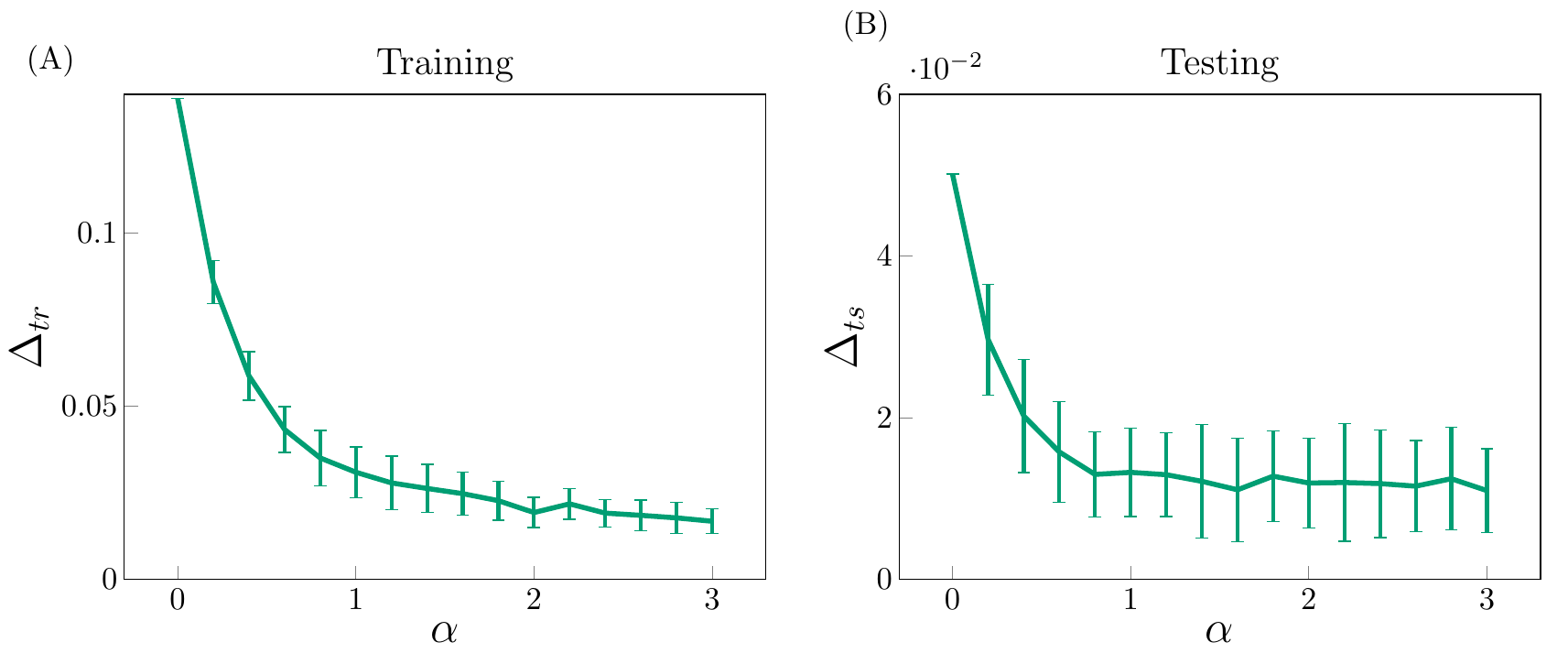}
			 \caption{Hindmarsh-Rose attractor. The training error (A) and testing error (B) vs $\alpha$. The parameter $\alpha$ controls the interval over which the random time-shifts are taken $[0,\bar{\tau}\alpha]$, where $\bar{\tau}$ is the characteristic time-scale of the Lorenz task and was found to be 0.46. {Error bars indicate the standard deviation over 50 iterations where each iteration corresponds to a different selection of the random time-shifts $\tau_i$, $\gamma=1.65$, $\epsilon=1$.}}\label{HRR}
		\end{figure}    
	\end{minipage}
\end{widetext}

The advantage of introducing random time-shifts is discussed in what follows for the case of three different `tasks', which we simply refer to as the chaotic Lorenz 96 system, the Lorenz system, and the Hindmarsh-Rose system. 

\subsection{Lorenz96 System Task}

\noindent The Lorenz96 chaotic system is modeled by the following set of equations,
\begin{equation}
    \begin{aligned} \label{L96}
        \dot x_i(t) & = (x_{i+1} - x_{i-2})x_{i-1} - x_i + F
    \end{aligned}
\end{equation}
where $i = 1,2,...,M$ for which we assume: $x_{-1} = x_{M-1}, x_0 = x_M, x_{M+1} = x_1$, $F=8$, $M=4$. 

We take the input signal $s(t)=x_1(t)$ and the training signal $g(t)=x_4(t)$. In the simulations the transient phase takes place from $0$ to $t_1 = 1000$, the training phase is from $t_1 = 1000$ to $t_2 = 1100$, and the testing phase is from $t_2 = 1100$ to $t_3 = 1200$.  

\noindent In Fig.\ \ref{L96R}, we plot the training ($\Delta_{tr}$) and testing error ($\Delta_{ts}$)  vs $\alpha$.
The figure shows a substantial improvement
 in both $\Delta_{tr}$ and $\Delta_{ts}$ when $\alpha$ is increased from $0$ to $5$. In particular, we see 
that the mere application of random time shifts to the readouts of a nonlinear reservoir computer leads to a reduction of both the training and testing error of roughly two orders of magnitude. 


\subsection{Lorenz System Task}

The Lorenz chaotic system is modeled by the following set of equations,
\begin{equation} \label{Lorenz system}
    \begin{aligned}
	    \dot x(t) & = c_1 (y(t)-x(t))\\
		\dot y(t) & = x(t)(c_2 -z(t))-y(t)\\
		\dot z(t) & = x(t)y(t) - c_3 z(t)
    \end{aligned}
\end{equation}
with $c_1 =10$, $c_2 =28$ and $c_3 = 8/3$. 

For this task, the $x(t)$ component is used as the input signal $s(t)$. The $y(t)$ component is used as the training signal. In simulation, the transient phase is set from $0$ to $t_1 = 600$, the training phase from $t_1 = 600$ to $t_2 = 610$ and the testing phase from $t_2 = 610$ to $t_3 = 615$.  

\noindent In Fig.\ \ref{LR}, we plot the training ($\Delta_{tr}$) and testing error ($\Delta_{ts}$)  vs $\alpha$. For all values of $\alpha>0$ we see a reduction in both the training and testing errors compared to the case that $\alpha=0$, which corresponds to the case in which time-shifts are not applied. 
In particular, we see that both the lowest training error and the lowest testing error are achieved at intermediate values of $\alpha$, with the best reduction  for $\alpha$ approximately equal to $0.2$. For larger values of $\alpha$ we see an increase of both the training and the testing error.

\subsection{Hindmarsh-Rose System Task}

\noindent The Hindmarsh-Rose chaotic system is modeled by the following equations,
\begin{equation} \label{Hindmarsh-Rose system}
    \begin{aligned}
	    \dot x (t) & = y (t) + \phi [x (t)] - z (t) + 1\\
	    \dot y (t) & = \psi [x (t)] -y (t)\\
	    \dot z (t) & = 5 \times 10^{-3}(4(x (t) + 8/5) - z (t))\\ 
	\end{aligned}
\end{equation}  
where,
\begin{equation*} 
	\begin{aligned}  
	    \phi [x(t)] &= -x^3 + 3x^2\\
	    \psi [x(t)] & = 1 - 5x^2.\\
    \end{aligned}
\end{equation*}

For this task, the $x(t)$ component is used as the input signal $s(t)$. The $y(t)$ component is used as the training  signal. In these simulations the transient phase takes place from $0$ to $t_1 = 1000$, the training phase from $t_1 = 1000$ to $t_2 = 1010$, and the testing phase  from $t_2 = 1010$ to $t_3 = 1015$. 

\noindent In Fig.\ \ref{HRR}, we plot the training ($\Delta_{tr}$) and testing error ($\Delta_{ts}$)  vs $\alpha$. For this case we see a substantial reduction in both the training error and testing error as we increase $\alpha$. The improvement seen in both $\Delta_{tr}$ and $\Delta_{ts}$ when $\alpha$ is increased from $0$ to $3$ is roughly of one order of magnitude.

\section{Optimization of Time-Shifts}

In this section, we describe a method to select the time-shifts that minimizes the training error. This method requires calculation of the time derivative of the reservoir response. We proceed under the assumption that all the time shifts are small. This assumption may be confirmed or not after computation of the optimized time-shifts. However, we decide to retain this assumption for the following two reasons: (i) it allows a simple solution to the optimization problem and (ii) even if the assumption is not verified by the optimized solution, we still hope that it will improve the RC performance with respect to either the case of no time-shifts or random time-shifts.  In what follows, we will test (ii) numerically for different choices of tasks and RC parameters. We will see that often times our strategy to optimize the time shifts overperforms random time shifts. 

After applying small time-shifts to the individual readouts of the RC, a first order Taylor expansion yields,
\begin{equation}
	r_i(t -\tau_i) \approx r_i(t) - \tau_i \dot{r}_i(t).
\end{equation}
Now the fit signal $\textbf{h}(t)$ in Eq.\ \eqref{eq:fit_delay} can be written as,
\begin{equation}
	\textbf{h} (t) = \sum_{i=1}^N \kappa_i r_i(t) + \kappa_{N+1} + \sum_i \lambda_i \dot{r}_i(t) , 
\end{equation}   
where $\lambda_i = - \kappa_i \tau_i$. In other words,
\begin{equation}
	\mathbf h = \left[\Omega_{r} \quad \Omega_{\dot{r}}\right]\begin{bmatrix} \pmb{\kappa} \\\pmb{\lambda}
	\end{bmatrix} = \Omega_{\textbf{opt}} \begin{bmatrix} \pmb{\kappa} \\\pmb{\lambda}
\end{bmatrix}, 
\end{equation}
where $\Omega_{r} \equiv \Omega$, $\pmb{\lambda} = \left[\lambda_1, \lambda_2,\cdots, \lambda_N\right]$ and 
\begin{equation}
	\Omega_{\dot{r}} =
	\left[\begin{array}{cccc}
		\dot{r}_1(1) & \dot{r}_2(1) & ... & \dot{r}_N(1) \\
		\dot{r}_1(2) & \dot{r}_2(2) & ... & \dot{r}_N(2) \\
		\vdots & \vdots & \vdots &  \vdots  \\
		\dot{r}_1(T) & \dot{r}_2(T) & ... & \dot{r}_N(T) \\
	\end{array}\right].
\end{equation}
The optimal coefficient vector can be found by solving the linear square fit with ridge regression,
\begin{equation}
	\begin{bmatrix} \pmb{\kappa}^* \\\pmb{\lambda}^*
	\end{bmatrix}  = 	\left(\Omega_{\text{opt}}^T\Omega_{\text{opt}} + \eta I\right)^{-1} \Omega_{\text{opt}}^T {\mathbf{g}}
\end{equation}
with optimized time-shifts  given by,
\begin{equation}
	\tau_i^* = - \lambda_i^*/\kappa_i^*, \text{ for } i = 1,2,\cdots,N.
\end{equation}
The fit signal with respect to the optimized time-shifts is,
\begin{equation}
	\mathbf h^* =  \Omega_{\text{shift}}^* \pmb{\kappa}^*,
\end{equation}
where $\Omega_{\text{shift}}^*$ is obtained by replacing $\tau_i$ by $\tau_i^*$ in Eq.\ \eqref{eq:Omega_delay} and the training error is computed by Eq.\ \eqref{eq:train}. The optimized time-shifts $\pmb{\tau}^*$ and $\pmb{\kappa}^*$ are then used to compute $\tilde\Omega_{\text{shift}}^*$ and the testing error. In all of our simulation in this article we set the ridge regression parameter $\eta$ to $10^{-6}$.

As we will see, though the optimization method is based on a first order approximation, it presents the main advantages that it is simple to compute numerically and it improves the accuracy and performance of the RC, compared to the case that no shifts are applied.

\subsection{Lorenz96 System Task}

\noindent Hereafter, we obtain the optimized time-shifts by the optimization method discussed in this section. The optimized time-shifts are applied to the individual nodes of the RC and the training and testing errors are computed.

Our best results are obtained for the case of the Lorenz96 System.
A comparison between application of random time-shifts and optimized time-shifts is presented in Fig.\ \ref{L96O}. In this case, optimized time-shifts perform much better than random time-shifts. In particular, for low values of $\gamma$ we see an improvement of many orders of magnitude, and a strong advantage of optimized time shifts is seen for all value of $\gamma$. Note also that random time-shifts still present a substantial improvement with respect to the case in which time-shifts are not used (compare Fig.\ \ref{L96R} and Fig.\ \ref{L96O}.)

\subsection{Lorenz System Task}


 In Fig.\ \ref{LO}, we plot the training error ($\Delta_{tr}$) and the testing error ($\Delta_{ts}$)  vs $\gamma$ for different RC configurations. 
 For $\gamma>3$ we see that the training error for the case of optimized time-shifts is lower by roughly one order of magnitude. However, in terms of testing error we do not see a benefit of using optimized time-shifts. We wish to emphasize that both random and optimized time-shifts present a substantial improvement with respect to the case in which time-shifts are not used (compare Fig.\ \ref{LR} and Fig.\ \ref{LO}.)

\subsection{Hindmarsh-Rose System Task}


\noindent A comparison between application of random time-shifts and optimized time-shifts is presented in Fig.\ \ref{HRO} for  the case of the Hindmarsh-Rose system. 
Fig.\ 
\ref{HRO} (A) is a plot of the training error ($\Delta_{tr}$) vs $\gamma$, showing that
for most values of $\gamma$ the RC accuracy is improved with optimized time-shifts. Fig.\ 
\ref{HRO} (B) is a plot of the testing error ($\Delta_{ts}$)  vs $\gamma$, showing that for most values of $\gamma$ in the range $1<\gamma<3$ the RC performance is improved with optimized time-shifts.

\section{Conclusion}

{This paper discussed the benefits associated with application of time shifts to the readouts of a reservoir computer. In all of our numerical experiments, we preliminarily optimize the RC parameters so to ensure we are working with a \textcolor{black}{well performing} reservoir. However, our results hold for generic RCs.}

For different `tasks', we \textcolor{black}{observe} that application of randomly chosen time shifts to the reservoir readouts leads to
	a substantial improvement in both accuracy (training error) and performance (testing error) \textcolor{black}{compared to the case in which time shifts are not used.} {The choice of random time shifts is consistent with the choice of a random topology for the connectivity of the RC network, which is commonly assumed in the literature. \textcolor{black}{We see} that the improvement observed is achieved independent of the particular selection of the time shifts.} 
	A further reduction in accuracy and performance is obtained when the time-shifts are computed by using a simple optimization approach. A case for which application of random and optimized time-shifts was particularly beneficial is that of the  Lorenz96 system (see Figs.\ \ref{L96R} and \ref{L96O}.) 



The method we use to optimize the time-shifts is very simple and at the same time, effective. 
Optimization methods such as Particle Swarm \cite{kennedy1995particle}, Simulated Annealing \cite{van1987simulated}, etc could be used to compute better approximations to the optimal time-shifts 
but these other optimization methods typically require much higher computational complexity  due to the large parameter space (in our case, a total of 100 time-shifts). On the other hand, the method we presented in this paper is  easily scalable.  

Our work may point out to a deeper connection with Taken's Embedding Theorem \cite{takens1981detecting}, which states that a chaotic attractor can be reconstructed from a single 'readout' function of the underlying dynamical system and linearly spaced delayed observations of this only readout function. Here we are using $N$ readouts and applying a different delay to each one of them. Exploring in more detail applications of Taken's Embedding Theorem to reservoir computers provides a promising direction for future investigation.
\color{black}

%
%

\begin{widetext}
	\begin{minipage}{\linewidth}
		\begin{figure}[H]
			\centering
			\includegraphics[width=0.99\textwidth]{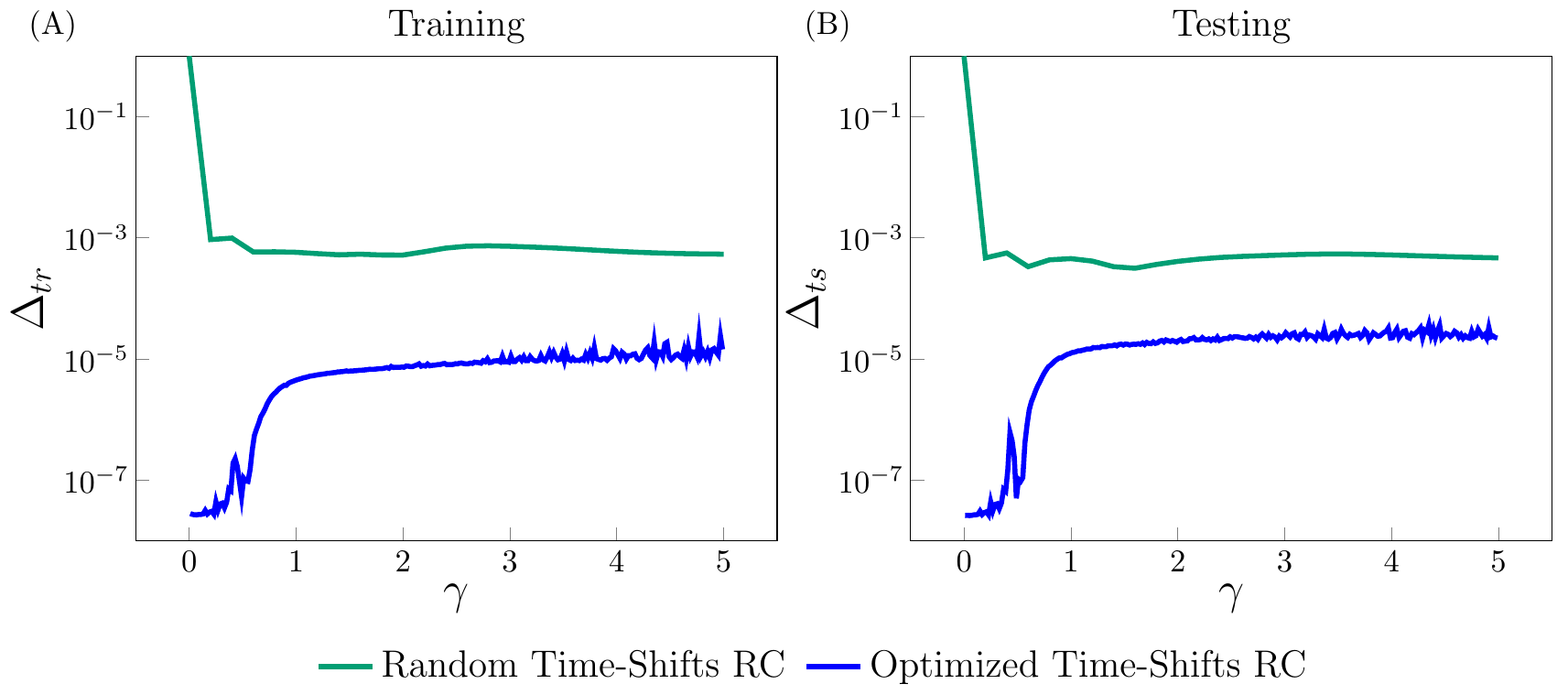}
			\caption{Lorenz96 system. The training error (A) and testing error (B) vs $\gamma$ for both the cases of: randomly drawn time shifts and optimized time shifts.   Here $\epsilon = 0.8$ and $\alpha = 4$.  } \label{L96O}
		\end{figure}    
	\end{minipage}
\end{widetext}

\begin{widetext}
	\begin{minipage}{\linewidth}
		\begin{figure}[H]
			\centering
			\includegraphics[width=0.99\textwidth]{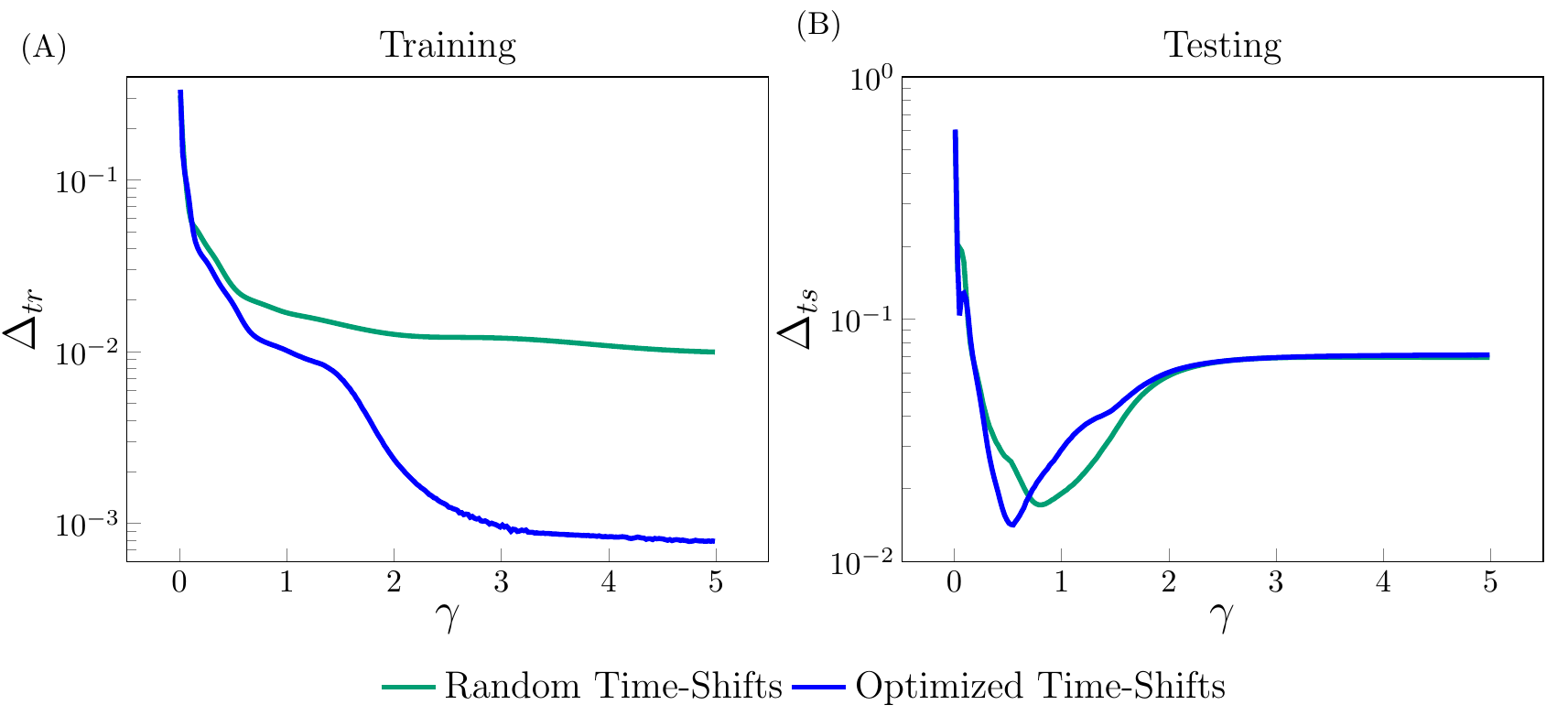}
			\caption{Lorenz attractor. The training error (A) and testing error (B) vs $\gamma$ are shown for the cases of randomly drawn times-shifts and optimized time-shifts. Here $\epsilon = 2$ and $\alpha = 0.25$. } \label{LO}
		\end{figure}    
	\end{minipage}
\end{widetext}
\begin{widetext}
	\begin{minipage}{\linewidth}
		\begin{figure}[H]
			\centering
			\includegraphics[width=0.99\textwidth]{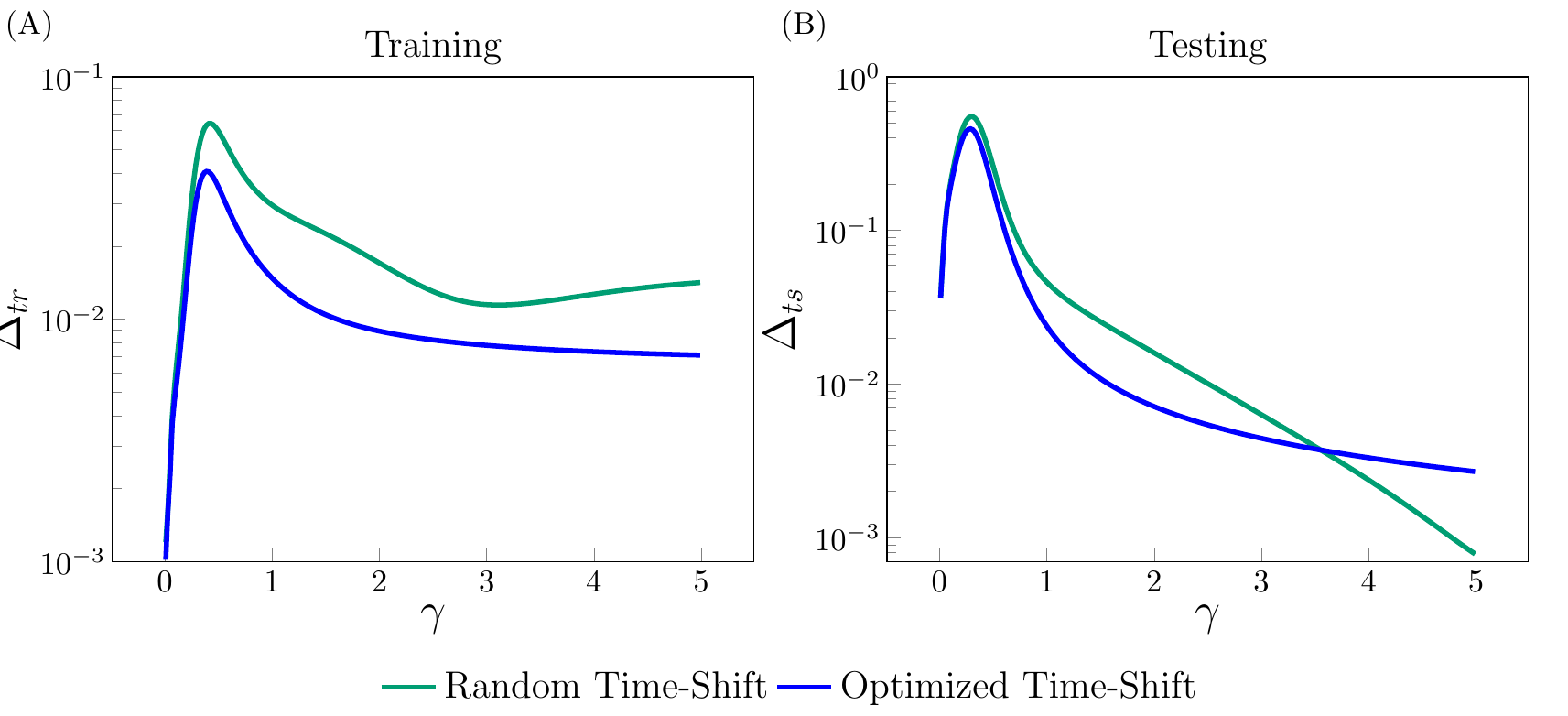}
			\caption{Hindmarsh-Rose attractor. The training error (A) and testing error (B) vs $\gamma$ for both  the cases of randomly drawn times-shifts and optimized time-shifts. Here $\epsilon = 1$ and $\alpha = 2.5$.} \label{HRO}
		\end{figure}    
	\end{minipage}
\end{widetext}

\section*{Acknowledgement}
The authors thank Lou Pecora and Tom Carroll for insightful conversations on the subject of Reservoir Computers. This research was supported by NIH (NIBIB) grant 1R21EB028489-01A1.

\section*{Author Declarations}
The authors have no conflicts to disclose.

\section*{Data Availability}
The data that supports the findings of this study are available within the article.

\end{document}